\documentclass{Interspeech2024}

\usepackage{hyperref}
\usepackage{multirow,multicol, makecell}
\usepackage{multirow}
\usepackage{multicol}
\usepackage{bm}
\usepackage{amssymb}
\usepackage{booktabs}
\usepackage{enumitem}
\usepackage{amsthm,amsmath,amssymb}
\usepackage{tabularx}
\usepackage{mathrsfs}
\usepackage{booktabs}
\usepackage{hyperref}

\interspeechcameraready 
\let\svthefootnote\thefootnote
\newcommand\blankfootnote[1]{%
  \let\thefootnote\relax\footnotetext{#1}%
  \let\thefootnote\svthefootnote%
}

\title{Improving Audio Codec-based Zero-Shot Text-to-Speech Synthesis with Multi-Modal Context and Large Language Model}

\name{Jinlong}{Xue}
\name{Yayue}{Deng}
\name{Yicheng}{Han}
\name{Yingming}{Gao}
\name[affiliation={*}]{Ya}{Li}

\address{
  Beijing University of Posts and Telecommunications, Beijing, China
  }
\email{\{jinlong\_xue, yayue.deng, adelacvgaoiro, yingming.gao, yli01\}@bupt.edu.cn}


\begin{document}

\maketitle

\blankfootnote{* Ya Li is the corresponding author.}

\begin{abstract}

    
    

Recent advances in large language models (LLMs) and development of audio codecs greatly propel the zero-shot TTS. They can synthesize personalized speech with only a 3-second speech of an unseen speaker as acoustic prompt. However, they only support short speech prompts and cannot leverage longer context information, as required in audiobook and conversational TTS scenarios. In this paper, we introduce a novel audio codec-based TTS model to adapt context features with multiple enhancements. Inspired by the success of Qformer, we propose a multi-modal context-enhanced Qformer (MMCE-Qformer) to utilize additional multi-modal context information. Besides, we adapt a pretrained LLM to leverage its understanding ability to predict semantic tokens, and use a SoundStorm to generate acoustic tokens thereby enhancing audio quality and speaker similarity. The extensive objective and subjective evaluations show that our proposed method outperforms baselines across various context TTS scenarios.

\end{abstract}

\begin{figure*}[t]
  \centering
  \includegraphics[width=\linewidth]{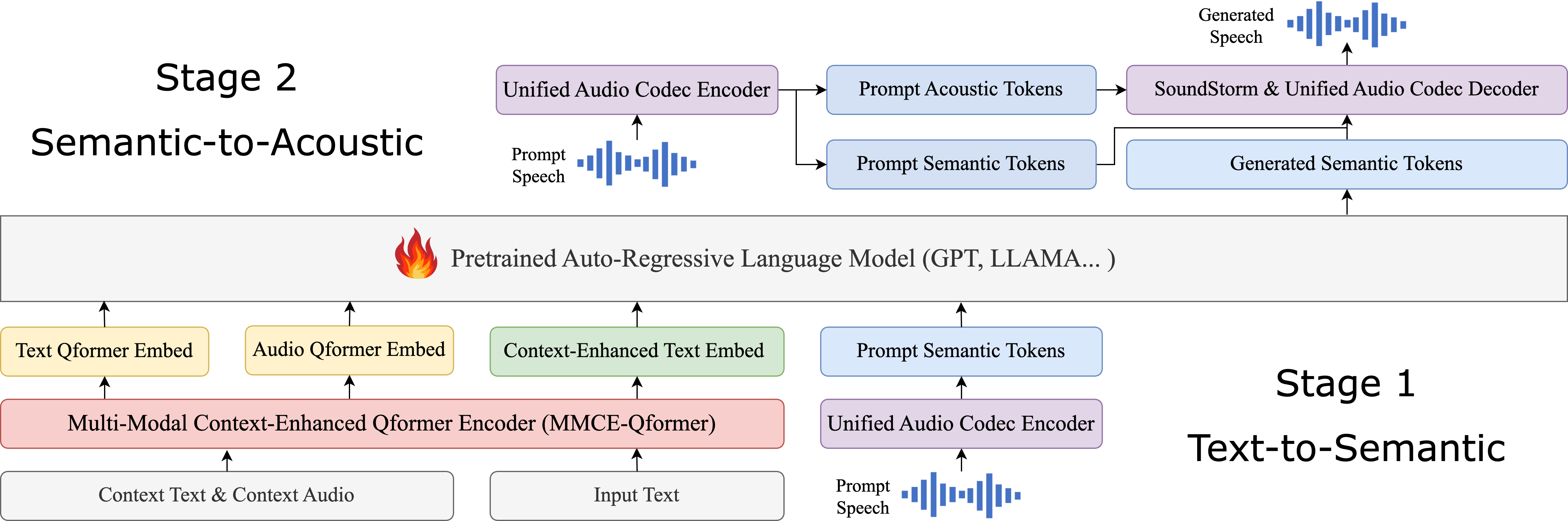}
  \caption{The overview of our proposed model.}
  \label{fig:overview}
  \vspace{-0.4cm}
\end{figure*}

\vspace{-0.2cm}
\section{Introduction}

Due to the use of deep learning architectures, text-to-speech (TTS) synthesis, which aims to generate natural speech from text, has seen tremendous improvements. Previous research has shown that neural TTS models~\cite{tacotron,fastspeech2,vits} can generate high-quality and natural speech, and models~\cite{cooper2020zerotts,casanova2021scglowtts,yourtts} can synthesize speech in multiple-speaker voices. However, these models still require a large amount of clean data for training and also need many speech utterances to fine-tune the models to adapt to new voices. Therefore, these models cannot fully exploit more real-world data but are limited to recorded data, and they have an inferior ability to clone unseen voices, which prevents the broad usage of zero-shot cloning.

Recent research \cite{valle,audiolm,speartts,tortoisetts} shows that zero-shot TTS has been greatly propelled by the use of neural audio codecs \cite{encodec,soundstream} which transits from continuous audio features to discrete tokens and the adoption of language model structures~\cite{gpt2,t5}. VALL-E~\cite{valle} adopts audio codec Encodec~\cite{encodec} with Residual Vector Quantization (RVQ) structure and treats TTS as a prompt-based language modeling task. It generates discrete audio codec outputs conditioned on text and audio codec prompts using a combination of autoregressive (AR) and non-autoregressive (NAR) transformer decoders. AudioLM~\cite{audiolm} employs hierarchical speech language models comprising text-to-semantic and semantic-to-acoustic stages using same decoder-only structure. It uses a semantic encoder to capture content information and an audio codec to model acoustic details. The above methods enable TTS models to train on large, diverse datasets from noisy multi-speaker environments and even real-world speech data, allowing the models to clone a speaker's voice with just a 3-second speech prompt. Compared to VALL-E using an NAR transformer decoder and AudioLM employing an AR semantic-to-acoustic model, SoundStorm~\cite{soundstorm} proposes a NAR confidence-based decoding scheme inspired by MaskGIT~\cite{maskgit} to generate acoustic tokens conditioned on semantic tokens.

However, the methods mentioned above have several limitations. Firstly, all the audio codec-based zero-shot TTS models depend solely on a speech prompt to provide the timbre and other acoustic information. However, they only support short speech samples for conditioning. This limitation is due to the high rate of acoustic codec tokens (75 Hz per RVQ level for Encodec) compared to the text token rate (usually around 10 Hz) in text generation~\cite{gpt2}. Therefore the acoustic token sequence is too long and 10-second speech has 750 tokens for one RVQ level, and 6K tokens for all 8 RVQ levels. As a result, although these models can clone a speaker's voice with just a short speech prompt, they cannot fully utilize longer additional information, like contextual information, as required in audiobook TTS~\cite{xue2022paratts,audiobookcontext,msstyletts} and conversational TTS~\cite{guo_CTTS,m2ctts,concss}. These studies find that enhancing TTS with contextual data can help improve prosody and even speaker similarity. However, how to model context information in current prompt-based TTS structure has not been fully explored.


Secondly, the first stage of these modern TTS plays an important role, such as the AR decoder in VALL-E and text-to-semantic AR model in AudioLM, since the generated tokens decide the most vital features in speech, like speaking style, prosody and emotion. However, these studies only train the first stage models from scratch and at phoneme level, while a pretrained LLM could enhance the understanding of input text content. Thirdly, the choices of the semantic encoders and audio codecs are sub-optimal. VALL-E relies solely on acoustic tokens struggling with content consistency due to the high acoustic rate and computational burden. Using both semantic encoder and audio codec, like AudioLM, leads to redundant overlapping features and repetitive computation.

To address the aforementioned limitations and enhance current audio codec-based zero-shot TTS models, we introduce a novel audio codec-based TTS model to adapt contextual information with several improvements. Inspired by the success of Qformer~\cite{blip2}, we introduce a multi-modal context-enhanced Qformer (MMCE-Qformer) encoder to fully utilize longer context information rather than relying solely on short speech prompts. We use learnable queries that act as a bottleneck to capture the most important global context, while the cross-attention mechanism helps extract relevant local context features for each text token. To further enhance the text-to-semantic process, we employ a pretrained LLM to inherit its understanding capacities. Besides, we utilize a unified audio codec to extract both semantic and acoustic tokens in a distillation manner. Additionally, we incorporate SoundStorm to enhance audio quality and speaker similarity. Our comprehensive experiments across various context scenarios demonstrate that our method surpasses existing audio codec-based zero-shot TTS models in naturalness, prosody, and speaker similarity through both subjective and objective evaluations. We also conduct ablation experiments to demonstrate the effectiveness of our model. Audio samples are available on the project page\footnote{https://happylittlecat2333.github.io/interspeech2024}. 


\begin{figure*}[t]
  \centering
  \includegraphics[width=0.9\linewidth]{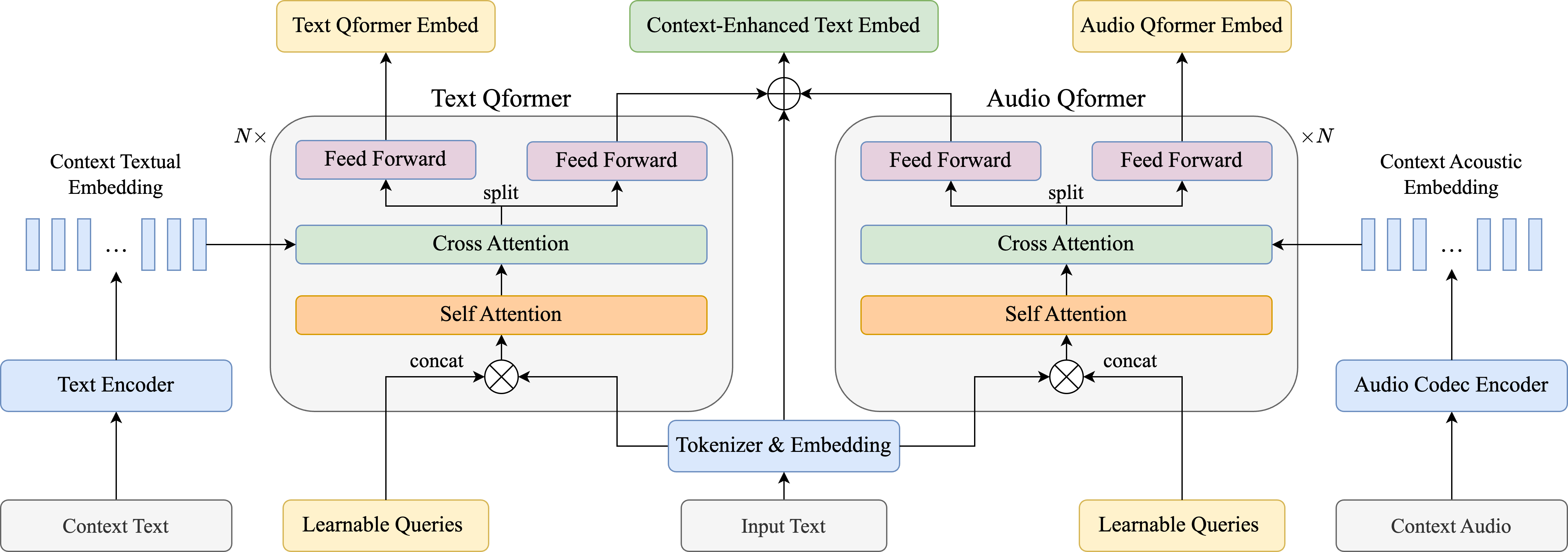}
  \caption{The structure of MMCE-Qformer encoder.}
  \label{fig:context_model}
  \vspace{-0.2cm}
\end{figure*}

\section{Methodology}

The overview of our proposed model is illustrated in Fig.\ref{fig:overview}. The whole model consists of four main components: a unified audio codec model for extracting semantic and acoustic tokens, a multi-modal context-enhanced Qformer encoder (MMCE-Qformer) to leverage multi-modal context information and enhance input text features, a text-to-semantic autoregressive model adapted from a pretrained LLM, and the SoundStorm model for the semantic-to-acoustic stage.

In the first stage, the MMCE-Qformer extracts text and audio Qformer embeddings as well as context-enhanced text embeddings from multi-modal contexts and input text. Then, the text-to-semantic model generates suitable semantic tokens based on the extracted features and prompt semantic tokens. We concatenate these inputs and feed them into the pretrained LLM, leveraging its understanding and in-context learning capabilities. In the second stage, we adopt SoundStorm to add acoustic features like timbre conditioned on generated semantic tokens and speech prompt to predict acoustic tokens. Finally, audio codec decoder converts output back to generated speech. The details for each part are described below.

\subsection{Unified Audio Codec}


We follow SpeechTokenizer~\cite{speechtokenizer} to extract semantic and acoustic tokens with one audio codec. It leverages semantic encoder HuBERT~\cite{hubert} as a teacher to distill the RVQ first level to hierarchically disentangle semantic and acoustic features in speech. Consequently, It unifies the semantic encoder and audio codec, where semantic tokens (RVQ first level) capture most content and paralinguistic features, and acoustic tokens (other levels) encode most acoustic features like speaker timbres.

\subsection{MMCE-Qformer Encoder}

The structure of our proposed MMCE-Qformer is illustrated in Fig.~\ref{fig:context_model}. The MMCE-Qformer is designed to extract multi-modal context embeddings and enhance the input text, by using arbitrarily long multi-modal context. It comprises two sub-models tailored to different modalities: the Text Qformer and the Audio Qformer. Both sub-models share an identical structure but are adapted to their respective modalities.

For the Audio Qformer, the input text is first tokenized and embedded by a pretrained language model to leverage its content understanding capabilities, instead of being converted into a phoneme sequence. To capture contextual audio features, we employ the same unified audio codec encoder but extract contextual acoustic embeddings instead of discrete tokens, aiming to preserve as much acoustic information as possible. Then, following the structure in Qformer~\cite{blip2}, the Audio Qformer consists of two transformer submodules that share the same attention layers: an audio transformer that interacts with contextual acoustic embeddings to extract the most important global acoustic features, and a text transformer that encodes the text input as well as guiding audio transformer to extract features most relevant to the text. We use fixed-length learnable vectors as queries to first interact with each other and additional text features through a self-attention block. A cross-attention block is introduced to obtain insights from the contextual acoustic embedding, with the concatenation of learnable queries and text embeddings serving as queries and the contextual acoustic embeddings as both keys and values. The output of the cross-attention block is split according to the lengths of the text embeddings and learnable queries and is processed through distinct feed-forward layers to produce the Audio Qformer Embeddings and Context-Enhanced Text Embeddings. We also add the text embeddings back residually to Context-Enhanced Text Embeddings. In this setup, learnable queries act as a bottleneck to capture the most crucial global acoustic context, while cross-attention and the residual addition of text embeddings aid in extracting relevant local acoustic context for each text token. Consequently, our approach effectively extracts both global and local acoustic context information to enhance zero-shot TTS. For the Text Qformer, it operates similarly but utilizes the text encoder RoBERTa~\cite{roberta} to extract textual context information and finally produce the Text Qformer Embeddings.

\subsection{Text-to-Semantic Generation}

To enhance current audio codec-based zero-shot TTS to utilize arbitrarily long context information, we compress the context into fixed-length embeddings as a multi-modal context prompt using the MMCE-Qformer introduced above. To leverage the strong in-context learning and content understanding ability of pretrained LLMs, we adapt a pretrained AR language model to generate semantic tokens guided by multi-modal context prompt, context-enhanced text embeddings, and semantic token prompt. Consequently, our text-to-semantic generation can leverage both global and local context information most relevant to the text from different modalities to enhance context understanding and rhythmic expression.

The detailed generation process is illustrated in Fig.\ref{fig:overview}, Text and Audio Qformer Embeddings $T$ and $A$, and Context-Enhanced Text Embeddings $x$ extracted from our MMCE-Qformer, along with prompt semantic tokens $\tilde{c}$ extracted from prompt speech are concatenated as input of the AR language model $\theta_{AR}$, and generate suitable semantic tokens $c$. Text and Audio Qformer Embeddings $T$ and $A$ are used for in-context learning and prompt semantic tokens guide the generation process to produce the desired speaker style. The AR generation process is formulated as follows:

\vspace{-1cm}
\begin{equation}
    p(c_{:} | x, \tilde{c}, T, A; \theta_{AR})=\prod_{t=0}^T p(c_{t} |c_{<t}, x, \tilde{c}, T, A ; \theta_{AR})
\end{equation}

\subsection{Semantic-to-Acoustic Generation}

Based on the unified audio codec, we use a modified SoundStorm model for the semantic-to-acoustic stage to predict acoustic tokens conditioned on semantic tokens and prompt speech, as shown in Fig.\ref{fig:overview}. SoundStorm is inspired by the confidence-based parallel decoding scheme of MaskGIT~\cite{maskgit} and adapted for RVQ token sequences. It utilizes a bidirectional attention-based Conformer~\cite{conformer} that is trained to predict masked audio tokens based on the summation of previous RVQ embeddings. During inference, given the conditioning (semantic tokens), SoundStorm starts with all acoustic tokens masked out except for the prompt and iteratively fills in the masked tokens parallelly, level by level. We follow the original decoding scheme, using (16, 1, 1, ..., 1) iterations for the RVQ levels, with 16 iterations at the first level and greedily choosing in the subsequent levels.

\section{Experiments}

\subsection{Training Setup}

We adopt the pretrained AR language model GPT-2~\cite{gpt2} and adapt it for text-semantic generation. Additionally, we utilize the GPT-2 tokenizer and its text embeddings to encode text inputs. Our proposed method is trained on the medium version of LibriLight~\cite{librilight}, which contains approximately 5K hours of unlabeled English speech from audiobooks. We employ Whisper~\cite{whisper} to transcribe the speech into text and segment it into utterances. Speech utterances shorter than 1 second or longer than 15 seconds are discarded, and all speech is converted to a 16kHz sampling rate. Finally, we have about 2M utterances in total. To augment the zero-shot TTS model with multi-modal context, we set the context length to 5 in the LibriLight dataset, leveraging its audiobooks for context-rich training. We select the preceding 5 utterances as context, incorporating both their text and speech. These are concatenated and fed into Text Qformer and Audio Qformer respectively.

We use the pretrained unified neural audio codec model SpeechTokenizer\footnote{https://github.com/ZhangXInFD/SpeechTokenizer} to extract both semantic and acoustic tokens and decode them into waveform. We also adopt its corresponding SoundStorm\footnote{https://github.com/ZhangXInFD/soundstorm-speechtokenizer} for semantic-to-acoustic stage. For Text Qformer and Audio Qformer, we use RoBERTa~\cite{roberta} as text encoder and adopt SpeechTokenizer as audio encoder. We follow Qformer~\cite{blip2} to use 32 queries with a dimension of 768, but using 2 layers. All models are trained for 300K iterations on one NVIDIA A6000 GPU with a batch size of 96, using the Adam optimizer with a learning rate of $1 \times 10^{-4}$ and bf16 precision.

\begin{table*}[t!]
\centering
\caption{Evaluation for zero-shot TTS among baselines and our models in both unseen LibriTTS test-clean and IEMOCAP datasets.}
\scalebox{0.79}{
\begin{tabular}{l|c|c|c|c|c|c|c|c|c|c|c|c}
\hline \hline 
 & \multicolumn{6}{c|}{LibriTTS test-clean} & \multicolumn{6}{c}{IEMOCAP} \\
\cline{2-13}
 & \multicolumn{4}{c|}{Objective} & \multicolumn{2}{c|}{Subjective}   & \multicolumn{4}{c|}{Objective} & \multicolumn{2}{c}{Subjective} \\
\hline        
Method      & Energy$\downarrow$ & F0$\downarrow$     & MCD$\downarrow$   & SECS$\uparrow$  & NMOS$\uparrow$  & SMOS$\uparrow$ & Energy$\downarrow$ & F0$\downarrow$     & MCD$\downarrow$    & SECS$\uparrow$  & NMOS$\uparrow$  & SMOS$\uparrow$ \\
\hline
Groundtruth & -      & -      & -     & -     & 4.900$\pm$0.075 & - & -      & -      & -      & -     & 4.717$\pm$0.047 & - \\
Reconstruct & 6.761 & 33.329 & 3.278 & 0.966 & 4.692$\pm$0.083 & 4.837$\pm$0.082 & 3.035  & 26.384 & 1.981  & 0.951 & 4.542$\pm$0.065 & 4.746$\pm$0.069 \\
\hline
VALL-E      & 19.809 & 65.749 & 8.670 & 0.710 & 3.283$\pm$0.130 & 3.371$\pm$0.124 & 21.843 & 83.561 & 8.855  & 0.649 & 3.188$\pm$0.121 & 3.212$\pm$0.114 \\
SoundStorm  & 18.080 & 59.957 & 7.974 & 0.795 & 3.412$\pm$0.135 & 3.608$\pm$0.136 & 22.000 & 77.812 & 8.579  & 0.682 & 3.254$\pm$0.128 & 3.508$\pm$0.124 \\
XTTS        & 20.467 & 56.190 & 7.442 & 0.754 & 3.671$\pm$0.087 & 3.533$\pm$0.093 & 24.359 & 72.408 & \textbf{8.451}  & 0.668 & 3.338$\pm$0.080 & 3.346$\pm$0.113 \\
\hline
\textbf{Proposed}    & \textbf{14.075} & \textbf{54.615} & \textbf{6.890} & \textbf{0.815} & \textbf{3.846}$\pm$\textbf{0.104} & \textbf{3.950}$\pm$\textbf{0.109} & \textbf{19.756} & 72.578 & 8.557  & \textbf{0.703} & \textbf{3.729}$\pm$\textbf{0.122} & \textbf{3.762}$\pm$\textbf{0.111} \\
w/o context & 14.681 & 54.943 & 7.099 & 0.803 & 3.796$\pm$0.083 & 3.775$\pm$0.090 & 20.367 & 73.435 & 8.721  & 0.688 & 3.438$\pm$0.138 & 3.608$\pm$0.133 \\
w/ text     & 14.588 & 54.724 & 7.046 & 0.806 & 3.708$\pm$0.097 & 3.805$\pm$0.104 & 20.455 & 73.023 & 8.677  & 0.690 & 3.508$\pm$0.123 & 3.654$\pm$0.113 \\
w/ audio    & 14.225 & 55.672 & 6.909 & 0.814 & 3.767$\pm$0.100 & 3.896$\pm$0.103 & 20.113 & \textbf{72.346} & 8.575  & 0.699 & 3.633$\pm$0.126 & 3.708$\pm$0.115 \\
\hline \hline
\end{tabular}
}
\label{tab:evaluation}
\vspace{-0.2cm}
\end{table*}

\vspace{-0.2cm}
\subsection{Evaluations}


To comprehensively evaluate the enhancement of our proposed method, we conduct evaluations in both audiobook and conversation scenarios using two unseen test datasets. We randomly select 20 chapters with totally 1249 utterances from unseen audiobooks dataset LibriTTS~\cite{libritts} (test-clean subset) for audiobook TTS evaluation. Besides, we use conversation dataset IEMOCAP~\cite{iemocap} for conversational TTS evaluation. IEMOCAP contains 151 dialogues and 7,433 utterances, amounting to nearly 12 hours of English dyadic conversations performed by ten professional actors. In each dyadic session, only two speakers participate. Since IEMOCAP is designed for emotion recognition task and is unsuitable for TTS due to noisy backgrounds and overlapping speeches, we follow~\cite{cmcu} to use pretrained denoiser model\footnote{https://github.com/facebookresearch/denoiser} to remove noise from all speech samples. We randomly select 16 dialogues including 1079 utterances for evaluation. We also use the preceding 5 utterances (text and speech) from the same speaker as the context in both audiobook and conversation TTS scenarios.

\vspace{-0.1cm}
\subsection{Compared Methods}

To show our model's performance, we compare it with the following models for zero-shot TTS. 
\begin{itemize}[topsep=2pt, partopsep=0pt, itemsep=0pt, parsep=0pt, leftmargin=*]
  \item \textbf{VALL-E}: an open-source implementation\footnote{https://github.com/0nutation/USLM} of VALL-E but use SpeechTokenizer as audio codec instead of Encodec. A 3-second prompt is used for zero-shot TTS.
  \item \textbf{SoundStorm}: only use AR stage of above VALL-E model and replace the NAR stage with the same SoundStorm model as ours. We also use the same 3-second prompt.
  \item \textbf{XTTS}: an open-source zero-shot TTS\footnote{https://huggingface.co/coqui/XTTS-v2} similar to TortoiseTTS~\cite{tortoisetts}, based on a discrete VAE model as audio codec and also use AR model, but it uses fixed-length vectors for speaker conditioning via Perceiver~\cite{perceiver} model.
  \item \textbf{Proposed}: our proposed method using multi-modal context for enhancement and also using the same 3-second prompt.
\end{itemize}


\vspace{-0.1cm}
\subsection{Objective Evaluation}

To evaluate the enhancements of our proposed model, including naturalness and speaker similarity in zero-shot synthesized speech, we calculate energy, F0, mel-cepstral distortion (MCD), and Speaker Encoder Cosine Similarity (SECS) as metrics for a thorough evaluation. We follow the previous method~\cite{yourtts} that uses speaker encoder Resemblyer\footnote{https://github.com/resemble-ai/Resemblyzer} to compute the SECS between groundtruth and generated speech. We apply dynamic time warping (DTW) to align the generated speech before calculating the differences in energy, F0, and MCD. The objective results are shown in Table~\ref{tab:evaluation}.

We find that SoundStorm significantly outperforms the VALL-E NAR model in all evaluation metrics. This highlights SoundStorm's superior ability to accurately predict the acoustic tokens conditioned on semantic tokens, thereby yielding higher quality speech. Since XTTS relies solely on global speaker vectors to guide the generation process, it exhibits less favorable outcomes in terms of speaker similarity and rhythm, where speaking style and prosody information are lost due to compression loss. We show that our proposed model surpasses all baselines in all evaluation metrics in LibriTTS and nearly all metrics in IEMOCAP test, particularly in speaker similarity and energy. Given that SoundStorm method adopts the same semantic-to-acoustic approach but utilizes a different text-to-semantic model, we attribute our success to the incorporation of multi-modal context and the adoption of a pretrained language model. We also conduct ablation studies on the impact of context from different modalities. Our experiments reveal that model without MMCE-Qformer yields inferior results. We also find that audio context, compared to text context, plays a more crucial role, significantly enhancing performance. This is likely because the audio modality encompasses richer information, such as timbre, prosody and emotion, and our Audio Qformer aids the text-to-semantic model by incorporating this additional information.

\vspace{-0.1cm}
\subsection{Subjective Evaluation}

To assess the naturalness and audio quality, as well as the speaking style and similarity of the generated speech, we conduct two subjective tests: a naturalness MOS (NMOS) test and a similarity MOS (SMOS) test. 20 randomly selected samples for each context scenario are evaluated by 10 listeners. Listeners rated the NMOS and SMOS on a scale of 1 to 5 in increments of 0.5. The subjective evaluation results are detailed in Table~\ref{tab:evaluation}. We find similar conclusions with our objective evaluation, showing that our proposed method outperforms all baselines and the adoption of MMCE-Qformer can encompass additional multi-modal context information for improvement.

\vspace{-0.1cm}
\section{Conclusion}

In this paper, we propose a novel audio codec-based TTS model that leverages long contextual information with multiple enhancements to improve both naturalness and speaker similarity. Inspired by the Qformer, we introduce the MMCE-Qformer to fully utilize multi-modal context features. This is achieved by employing learnable queries that act as a bottleneck to capture important global context, along with cross-attention layers that extract relevant local context features for each text token. Additionally, we adapt a pretrained LLM to enhance text-to-semantic generation and employ SoundStorm to improve audio quality and speaker similarity. Our evaluation across various contextual TTS scenarios demonstrates that our proposed method outperforms all baseline models. 

\vspace{-0.1cm}
\section{Acknowledgements}
The work was supported by the National Natural Science Foundation of China (NSFC) (No. 62271083),  the Key Project of the National Language Commission (No. ZDI145-81), and the Fundamental Research Funds for the Central Universities (No. 2023RC73, 2023RC13).


\newpage

\bibliographystyle{IEEEtran}
\bibliography{mybib}

\begin{thebibliography}{10}
\providecommand{\url}[1]{#1}
\csname url@samestyle\endcsname
\providecommand{\newblock}{\relax}
\providecommand{\bibinfo}[2]{#2}
\providecommand{\BIBentrySTDinterwordspacing}{\spaceskip=0pt\relax}
\providecommand{\BIBentryALTinterwordstretchfactor}{4}
\providecommand{\BIBentryALTinterwordspacing}{\spaceskip=\fontdimen2\font plus
\BIBentryALTinterwordstretchfactor\fontdimen3\font minus \fontdimen4\font\relax}
\providecommand{\BIBforeignlanguage}[2]{{%
\expandafter\ifx\csname l@#1\endcsname\relax
\typeout{** WARNING: IEEEtran.bst: No hyphenation pattern has been}%
\typeout{** loaded for the language `#1'. Using the pattern for}%
\typeout{** the default language instead.}%
\else
\language=\csname l@#1\endcsname
\fi
#2}}
\providecommand{\BIBdecl}{\relax}
\BIBdecl

\bibitem{tacotron}
Y.~Wang, R.~Skerry-Ryan, D.~Stanton, Y.~Wu, R.~J. Weiss, N.~Jaitly, Z.~Yang, Y.~Xiao, Z.~Chen, S.~Bengio \emph{et~al.}, ``Tacotron: Towards end-to-end speech synthesis,'' \emph{arXiv preprint arXiv:1703.10135}, 2017.

\bibitem{fastspeech2}
Y.~Ren, C.~Hu, X.~Tan, T.~Qin, S.~Zhao, Z.~Zhao, and T.-Y. Liu, ``Fastspeech 2: Fast and high-quality end-to-end text to speech,'' \emph{arXiv preprint arXiv:2006.04558}, 2020.

\bibitem{vits}
J.~Kim, J.~Kong, and J.~Son, ``Conditional variational autoencoder with adversarial learning for end-to-end text-to-speech,'' in \emph{International Conference on Machine Learning}.\hskip 1em plus 0.5em minus 0.4em\relax PMLR, 2021, pp. 5530--5540.

\bibitem{cooper2020zerotts}
E.~Cooper, C.-I. Lai, Y.~Yasuda, F.~Fang, X.~Wang, N.~Chen, and J.~Yamagishi, ``Zero-shot multi-speaker text-to-speech with state-of-the-art neural speaker embeddings,'' in \emph{ICASSP 2020-2020 IEEE International Conference on Acoustics, Speech and Signal Processing (ICASSP)}.\hskip 1em plus 0.5em minus 0.4em\relax IEEE, 2020, pp. 6184--6188.

\bibitem{casanova2021scglowtts}
E.~Casanova, C.~Shulby, E.~G{\"o}lge, N.~M. M{\"u}ller, F.~S. de~Oliveira, A.~C. Junior, A.~d.~S. Soares, S.~M. Aluisio, and M.~A. Ponti, ``Sc-glowtts: an efficient zero-shot multi-speaker text-to-speech model,'' \emph{arXiv preprint arXiv:2104.05557}, 2021.

\bibitem{yourtts}
E.~Casanova, J.~Weber, C.~D. Shulby, A.~C. Junior, E.~G{\"o}lge, and M.~A. Ponti, ``Yourtts: Towards zero-shot multi-speaker tts and zero-shot voice conversion for everyone,'' in \emph{International Conference on Machine Learning}.\hskip 1em plus 0.5em minus 0.4em\relax PMLR, 2022, pp. 2709--2720.

\bibitem{valle}
C.~Wang, S.~Chen, Y.~Wu, Z.~Zhang, L.~Zhou, S.~Liu, Z.~Chen, Y.~Liu, H.~Wang, J.~Li \emph{et~al.}, ``Neural codec language models are zero-shot text to speech synthesizers,'' \emph{arXiv preprint arXiv:2301.02111}, 2023.

\bibitem{audiolm}
Z.~Borsos, R.~Marinier, D.~Vincent, E.~Kharitonov, O.~Pietquin, M.~Sharifi, D.~Roblek, O.~Teboul, D.~Grangier, M.~Tagliasacchi \emph{et~al.}, ``Audiolm: a language modeling approach to audio generation,'' \emph{IEEE/ACM Transactions on Audio, Speech, and Language Processing}, 2023.

\bibitem{speartts}
E.~Kharitonov, D.~Vincent, Z.~Borsos, R.~Marinier, S.~Girgin, O.~Pietquin, M.~Sharifi, M.~Tagliasacchi, and N.~Zeghidour, ``Speak, read and prompt: High-fidelity text-to-speech with minimal supervision,'' \emph{arXiv preprint arXiv:2302.03540}, 2023.

\bibitem{tortoisetts}
J.~Betker, ``Better speech synthesis through scaling,'' \emph{arXiv preprint arXiv:2305.07243}, 2023.

\bibitem{encodec}
A.~D{\'e}fossez, J.~Copet, G.~Synnaeve, and Y.~Adi, ``High fidelity neural audio compression,'' \emph{arXiv preprint arXiv:2210.13438}, 2022.

\bibitem{soundstream}
N.~Zeghidour, A.~Luebs, A.~Omran, J.~Skoglund, and M.~Tagliasacchi, ``Soundstream: An end-to-end neural audio codec,'' \emph{IEEE/ACM Transactions on Audio, Speech, and Language Processing}, vol.~30, pp. 495--507, 2021.

\bibitem{gpt2}
A.~Radford, J.~Wu, R.~Child, D.~Luan, D.~Amodei, I.~Sutskever \emph{et~al.}, ``Language models are unsupervised multitask learners,'' \emph{OpenAI blog}, vol.~1, no.~8, p.~9, 2019.

\bibitem{t5}
C.~Raffel, N.~Shazeer, A.~Roberts, K.~Lee, S.~Narang, M.~Matena, Y.~Zhou, W.~Li, and P.~J. Liu, ``Exploring the limits of transfer learning with a unified text-to-text transformer,'' \emph{The Journal of Machine Learning Research}, vol.~21, no.~1, pp. 5485--5551, 2020.

\bibitem{soundstorm}
Z.~Borsos, M.~Sharifi, D.~Vincent, E.~Kharitonov, N.~Zeghidour, and M.~Tagliasacchi, ``Soundstorm: Efficient parallel audio generation,'' \emph{arXiv preprint arXiv:2305.09636}, 2023.

\bibitem{maskgit}
H.~Chang, H.~Zhang, L.~Jiang, C.~Liu, and W.~T. Freeman, ``Maskgit: Masked generative image transformer,'' in \emph{Proceedings of the IEEE/CVF Conference on Computer Vision and Pattern Recognition}, 2022, pp. 11\,315--11\,325.

\bibitem{xue2022paratts}
L.~Xue, F.~K. Soong, S.~Zhang, and L.~Xie, ``Paratts: Learning linguistic and prosodic cross-sentence information in paragraph-based tts,'' \emph{IEEE/ACM Transactions on Audio, Speech, and Language Processing}, vol.~30, pp. 2854--2864, 2022.

\bibitem{audiobookcontext}
D.~Xin, S.~Adavanne, F.~Ang, A.~Kulkarni, S.~Takamichi, and H.~Saruwatari, ``Improving speech prosody of audiobook text-to-speech synthesis with acoustic and textual contexts,'' in \emph{ICASSP 2023-2023 IEEE International Conference on Acoustics, Speech and Signal Processing (ICASSP)}.\hskip 1em plus 0.5em minus 0.4em\relax IEEE, 2023, pp. 1--5.

\bibitem{msstyletts}
S.~Lei, Y.~Zhou, L.~Chen, Z.~Wu, X.~Wu, S.~Kang, and H.~Meng, ``Msstyletts: Multi-scale style modeling with hierarchical context information for expressive speech synthesis,'' \emph{IEEE/ACM Transactions on Audio, Speech, and Language Processing}, 2023.

\bibitem{guo_CTTS}
H.~Guo, S.~Zhang, F.~K. Soong, L.~He, and L.~Xie, ``Conversational end-to-end tts for voice agents,'' in \emph{2021 IEEE Spoken Language Technology Workshop (SLT)}.\hskip 1em plus 0.5em minus 0.4em\relax IEEE, 2021, pp. 403--409.

\bibitem{m2ctts}
J.~Xue, Y.~Deng, F.~Wang, Y.~Li, Y.~Gao, J.~Tao, J.~Sun, and J.~Liang, ``M2-ctts: End-to-end multi-scale multi-modal conversational text-to-speech synthesis,'' in \emph{ICASSP 2023-2023 IEEE International Conference on Acoustics, Speech and Signal Processing (ICASSP)}.\hskip 1em plus 0.5em minus 0.4em\relax IEEE, 2023, pp. 1--5.

\bibitem{concss}
Y.~Deng, J.~Xue, Y.~Jia, Q.~Li, Y.~Han, F.~Wang, Y.~Gao, D.~Ke, and Y.~Li, ``Concss: Contrastive-based context comprehension for dialogue-appropriate prosody in conversational speech synthesis,'' \emph{arXiv preprint arXiv:2312.10358}, 2023.

\bibitem{blip2}
J.~Li, D.~Li, S.~Savarese, and S.~Hoi, ``Blip-2: Bootstrapping language-image pre-training with frozen image encoders and large language models,'' \emph{arXiv preprint arXiv:2301.12597}, 2023.

\bibitem{speechtokenizer}
X.~Zhang, D.~Zhang, S.~Li, Y.~Zhou, and X.~Qiu, ``Speechtokenizer: Unified speech tokenizer for speech large language models,'' \emph{arXiv preprint arXiv:2308.16692}, 2023.

\bibitem{hubert}
W.-N. Hsu, B.~Bolte, Y.-H.~H. Tsai, K.~Lakhotia, R.~Salakhutdinov, and A.~Mohamed, ``Hubert: Self-supervised speech representation learning by masked prediction of hidden units,'' \emph{IEEE/ACM Transactions on Audio, Speech, and Language Processing}, vol.~29, pp. 3451--3460, 2021.

\bibitem{roberta}
Y.~Liu, M.~Ott, N.~Goyal, J.~Du, M.~Joshi, D.~Chen, O.~Levy, M.~Lewis, L.~Zettlemoyer, and V.~Stoyanov, ``Roberta: A robustly optimized bert pretraining approach,'' \emph{arXiv preprint arXiv:1907.11692}, 2019.

\bibitem{conformer}
A.~Gulati, J.~Qin, C.-C. Chiu, N.~Parmar, Y.~Zhang, J.~Yu, W.~Han, S.~Wang, Z.~Zhang, Y.~Wu \emph{et~al.}, ``Conformer: Convolution-augmented transformer for speech recognition,'' \emph{arXiv preprint arXiv:2005.08100}, 2020.

\bibitem{librilight}
J.~Kahn, M.~Rivi{\`e}re, W.~Zheng, E.~Kharitonov, Q.~Xu, P.-E. Mazar{\'e}, J.~Karadayi, V.~Liptchinsky, R.~Collobert, C.~Fuegen \emph{et~al.}, ``Libri-light: A benchmark for asr with limited or no supervision,'' in \emph{ICASSP 2020-2020 IEEE International Conference on Acoustics, Speech and Signal Processing (ICASSP)}.\hskip 1em plus 0.5em minus 0.4em\relax IEEE, 2020, pp. 7669--7673.

\bibitem{whisper}
A.~Radford, J.~W. Kim, T.~Xu, G.~Brockman, C.~McLeavey, and I.~Sutskever, ``Robust speech recognition via large-scale weak supervision,'' in \emph{International Conference on Machine Learning}.\hskip 1em plus 0.5em minus 0.4em\relax PMLR, 2023, pp. 28\,492--28\,518.

\bibitem{libritts}
H.~Zen, V.~Dang, R.~Clark, Y.~Zhang, R.~J. Weiss, Y.~Jia, Z.~Chen, and Y.~Wu, ``Libritts: A corpus derived from librispeech for text-to-speech,'' \emph{arXiv preprint arXiv:1904.02882}, 2019.

\bibitem{iemocap}
C.~Busso, M.~Bulut, C.-C. Lee, A.~Kazemzadeh, E.~Mower, S.~Kim, J.~N. Chang, S.~Lee, and S.~S. Narayanan, ``Iemocap: Interactive emotional dyadic motion capture database,'' \emph{Language resources and evaluation}, vol.~42, pp. 335--359, 2008.

\bibitem{cmcu}
Y.~Deng, J.~Xue, F.~Wang, Y.~Gao, and Y.~Li, ``Cmcu-css: Enhancing naturalness via commonsense-based multi-modal context understanding in conversational speech synthesis,'' in \emph{Proceedings of the 31st ACM International Conference on Multimedia}, 2023, pp. 6081--6089.

\bibitem{perceiver}
A.~Jaegle, F.~Gimeno, A.~Brock, O.~Vinyals, A.~Zisserman, and J.~Carreira, ``Perceiver: General perception with iterative attention,'' in \emph{International conference on machine learning}.\hskip 1em plus 0.5em minus 0.4em\relax PMLR, 2021, pp. 4651--4664.

\end{thebibliography}

\end{document}